# Scaled demagnetization models: susceptibility, resonant and relaxation frequency prediction compared to magnetic composite measurements

Dr. Rick L. Moore, Georgia Tech Research Institute, Georgia Institute of Technology, Atlanta Georgia 30332, USA, rick.moore@gtri.gatech.edu

*Abstract* — Maxwell-Garnett Theory (MGT), Bruggeman Effective Medium Theory (BEMT), Coherent Model Approximation (CMA), Scaled Effective Medium Theory (ScEMT) and models of Schlomann (E. Schlomann, Phys. Rev. B, 182, 7, 632 (10 June 1969) and E. Schlomann, Conf. on Mag. and Mag. Materials, AIEE Spec. Publ. T-91, 600 (1956)) are applied to predict magnetic susceptibility, resonant and relaxation frequency in polymer-magnetic particle composites. Particulates had aspect ratios near unity; bulk low frequency susceptibilities ranging from approximately 5 to 4000 and particle volume fractions between 1 and 100%. Previous publications demonstrated that ScEMT improved the prediction of DC susceptibility as compared to classical models. This paper first modifies BEMT and ScEMT for volume fractions below about 10%. A ScEMT based model of composite resonant frequency is presented and compared to MGT, CMA models and measurement. Model and measurement comparisons of resonant frequency are followed by model-measurement comparisons of relaxation frequency. CMA, MGT, models of Schlomann, and volumetric scaled modifications of Schlomann are tested in the relaxation frequency study. The paper emphasizes the broad application of the models and therefore composite data for a wide range of particulate chemistries are presented ScEMT predictions of susceptibility and resonant frequency continue to show reasonable agreement with measurement and represent improvement over the CMA and MGT models. ScEMT modifications to Schlomann show overall best agreement with relaxation frequency measurement. CMA and MGT are most accurate for modest susceptibility (~ < 100) while ScEMT modified Schlomann models are most accurate for large susceptibility.

# Introduction

Composites formed as mixtures of magnetic and nonmagnetic materials are applied in a wide range of modern RF technologies. New ferrite compositions, many developed for applications in bio-magnetics, and new particulate geometries are in constant development and these provide new candidates for making improved composites. Electromagnetic interference (EMI) suppression and antenna substrates are two applications; additional applications are listed in [1] [2]. Composites with desired electromagnetic constitutive parameters are often developed by experimentation using an iterative process of formulation-measurement-formulation. This procedure introduces a latency in the design process of making optimized material formulations and experimental approaches can investigate only a limited parameter space. This lag can be reduced and parameter spaces fully investigated by using accurate predictive models of electromagnetic parameters.

This paper is an extension of the previously developed and experimentally tested volumetrically scaled effective medium theory (ScEMT [3] [4]). ScEMT demonstrated significant improvements in the prediction of low frequency composite magnetic susceptibility as compared to other effective medium models and early results showed promise for predicting composite resonant frequency. In this work, the predictive accuracies of ScEMT, CMA, MGT [5] [6]and Schlomann[7] [8] models are stressed by comparing to measurement for a very wide range of magnetic particulates, composite volume fractions and frequency.

This paper begins with a short review of the basis for ScEMT and analyzes a low volume fraction correction for symmetric microstructure models. The correction is for magnetic volume fractions below about 10%.

Though the correction does bring measurement, dispersive and symmetric composite model predictions closer, it does not completely reconcile the discrepancies in the data sets. The corrected model is compared for low and high susceptibility particulate composites.

Next, the ScEMT contribution to prediction of resonant frequency is reviewed and compared to a model selection [9] that builds on MGT[5] and CMA[6]. All models are compared to a broad selection of resonant frequency measured data and thus expands the analysis of [4]

The paper concludes with studies using CMA, MGT and Schlomann [7] [8]models that predict the relaxation frequency in magnetic composites. CMA and MGT are compared to analysis of the 1969 reference1 [8] and ScEMT modifications of [8]. Copies of Schlomann's 1956 paper [7] are somewhat difficult to obtain however Lax and Button, Section 10.3 [10] and R. Krishnan reference [11] each present data supporting both 1956 and 1969 papers. After volumetric scales taken from ScEMT are added to the 1969 analysis, a significant improvement in predictive accuracy is obtained; as compared to measurement, MGT and CMA. Improvement is demonstrated using a wide range of magnetic particulate and composite measured relaxation frequencies.

## ScEMT Review

ScEMT is built upon the Bruggeman effective medium theory, BEMT [12]. The BEMT formulation was modified to include a nonlinear dependence of magnetic demagnetization on volume fraction. The nonlinear approached is supported by publications by Chevalier and Le Floc'h [13], M. Anhalt, et.al. [14], C. Alvarez and S. H. L. Klapp [15]. J. L. Mattei and collaborators, [16] [17] [18], also elucidate the need for nonlinearity to predict susceptibility.

Magnetic constitutive equations for the CMA, BEMT, MGT and ScEMT are found below for two-phase composites. CMA and MGT equations are simplified from the standard format by specializing them for nonmagnetic matrices. Subscripts e, p and m indicate composite, particulate and matrix parameters respectively with $P_p$ being volume fraction of the particulate, and $\chi_p$ as the particulate's magnetic susceptibility. $A_B$ is demagnetization which is 1/3 for spherical particulates like those assumed in this paper.

$$CMA: \chi_e = \frac{P_p^{1/3}\chi_p}{1+(1-P_p^{1/3})\chi_p} = \frac{P_p\chi_p}{P_p^{1/3}(1+\chi_p)-P_p^{1/3}\chi_p} \quad (1)$$

$$MGT: \chi_e = \frac{3P_p\chi_p}{(3+\chi_p)-P_p\chi_p} \quad (2)$$

$$BEMT: P_p\frac{\mu_e-\mu_p}{\mu_e+(\mu_p-\mu_e)A_B} + (1-P_p)\frac{\mu_e-\mu_m}{\mu_e+(\mu_m-\mu_e)A_B} = 0 \quad (3)$$

The BEMT equation for spherical particulates describes a binary system whose microstructure changes from differentiated to symmetric near a volume fraction of 1/3, the percolation threshold $P_C$. J. P. Clerc, et.al [19] notes that when $A_B = 1/3$, $\mu_e$ in the denominator of Equation 3 becomes multiplied by the factor 2. In general that factor is $d - 1$ where $d$ is the system dimensionality ($d - 1 = 2$ in 3D and $d - 1 = 1$ in 2D) and the factor represents demagnetization in the composite. System dimensionality is connected to the percolation threshold by $P_C = d^{-1}$.

Demagnetization in a composite has a geometrical and internal component. If particulate and matrix have different susceptibilities, magnetic poles are formed at particulate-matrix interfaces and produce an opposing internal demagnetization field within particulates. Formation of poles and demagnetization also changes with particulate shape and magnetic coupling or chaining of magnetic particulates. Thus, demagnetization is a function of particulate susceptibility, particulate cluster sizes and shapes, and the media susceptibility that surrounds the particulate. There is an inherent nonlinearity in the system description since demagnetization is itself a function of composite susceptibility, which changes with particulate volume fraction.

Application of percolation theory [19] suggests insight into nonlinear geometrical relations among demagnetization, volume fraction, dimensionality and percolation threshold. In a composite, the largest cluster size $S(a, P_p, P_c)$ is approximately

$$S(a, P_p, P_c) = a(P_p - P_c)^{-\gamma} \tag{4}$$

where $a$ is the characteristic dimension of a single particle and $\gamma$ is a non-integer exponent that changes with dimension of the composite. Since demagnetization is a function of magnetic cluster size and shape $S(a, P_p, P_c)$; Equation 4 suggests that demagnetization will have a nonlinear dependence volume fraction.

Numerical simulation, and cluster experiments of [3] were used to identify a nonlinear function, $A_p$, that relates demagnetization, percolation threshold and volume fraction. $A_p$ replaces the classical BEMT demagnetization constant, $d - 1 = \frac{1-P_C}{P_C}$, by the ratio $A_p = \frac{1-A_C}{A_C}$, i.e.

$$ScEMT: P_p \frac{\mu_e - \mu_p}{A_p \mu_e + \mu_p} + (1 - P_p) \frac{\mu_e - \mu_m}{A_p \mu_e + \mu_m} = 0 \tag{5}$$

where

$$A_c = A_0\{1. - A_1(1 - P_p)^\gamma\} \text{ and } A_p = \left(\frac{1-A_C}{A_C}\right). \tag{6}$$

These equations incorporate the nonlinear relationship between volume fraction or cluster shape and percolation threshold. Magnetic particulate percolation (i.e. clustering) is not random but is biased by internal magnetic fields that couple particulate poles. Strong coupling between particulates increases probability of particulate chaining to minimize the overall system energy and is believed to contribute to ScEMT accuracy with large susceptibility particulates.

The free parameters, $A_0$, $A_1$ and $\gamma$ were determined by fitting the ScEMT Equation 5 to measurement of non-dispersive, low frequency (< 10 MHz) susceptibilities of composites made from 10 - 40 $\mu m$, multi-domain Ni $_{0.31}$ Zn $_{0.58}$ Cu$_{0.08}$ F$_{2.03}$ O$_4$. Samples were made with ferrite volume concentrations of 18, 20, 21, 30, 32, 45, 57 and 64 %, [3] [4]. A 100% dense control sample was also measured and found DC susceptibility of 863. The 100% data supplied the base input for model testing. Composite and solid samples were cut to size for a 7 mm coaxial transmission line test fixture. Complex reflection and transmission measurements were measured and from those permittivity and permeability were calculated. Those data were then fit to a Landau-Lifshitz-Gilbert resonance equation. DC susceptibility, resonant frequency and relaxation parameters could be inferred from that fit. Details of measurement and procedures can be found in [20] along with measured data of the 100% dense ferrite.

The fitted values for the ScEMT scaling function are near unity with numerical values of: $A_0 = 0.975$, $A_1 = 0.923$ and $\gamma = 1.210$, when rounded to the nearest thousandth. Ideally coefficients would be equal to satisfy the ScEMT boundary prediction at unity volume fraction. At that fraction composite susceptibility and particulate susceptibility must be equal. This is requires that $\left(\frac{1-A_C}{A_C}\right)$ goes too zero at unity fraction. In the ScEMT model with fitted parameters, $A_0$, $A_1$ and $\gamma$, the ratio approaches 0.025 as concentration goes to unity. At zero volume fraction, the ratio approaches 12.32 whereas it should go to infinity. The numeric differences indicate that small errors should be expected at high and low volume fractions. However, as [4] [3] demonstrates the fitted parameter values significantly improved model fit to published measurements for a wide range of magnetic particulate in nonmagnetic matrices for particulate volume fractions above 10 %and particulate susceptibilities of tens to thousands. In discussions of bandwidth, we do find that by forcing $A_0$ and $A_1$ to be equal, fit to measured relaxation bandwidths is improved.

### Small Volume Fraction Correction to BEMT and ScEMT

Due to differences in composite microstructures, predictions of BEMT and ScEMT should be expected to deviate from measurement at volume fractions approaching zero or unity. Each model assumes composites with symmetric microstructure; i.e. under exchange of particulate, matrix and fractional content, the EMT require equivalent results. However, at very low or very high volume fractions the composite microstructure should be dispersive or differentiated. Microstructure geometries are illustrated in Figure 1. Particulates are nearly isolated in the dispersive case and therefore scattering theories like MGT should apply. Differences in symmetric vs. differentiated microstructures are discussed in [21] and [22]. Small volume fraction expansions of MGT, ScEMT and BEMT produce analytical solutions for a range of particulate susceptibilities and expansions are found in [21]. The expansion for ScEMT applies the BEMT expansion but where the constant demagnetization factor of 2 is replaced by $A_p$, Equation 6.

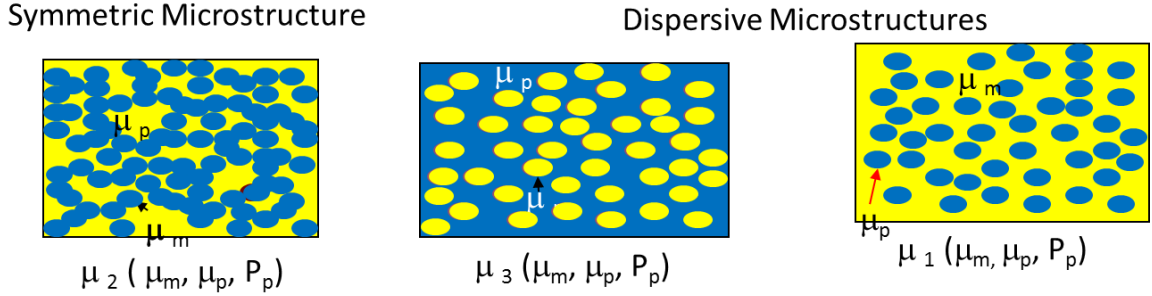

Figure 1 The above are graphical representations of symmetric microstructure (for BEMT and ScEMT) and dispersive (for MGT) microstructures.

Equations 7-9 respectively show MGT, BEMT and ScEMT model expansions predicting the ratio of effective permeability to host permeability in powers of volume fraction to second order. Subscripts MGT, BEMT and ScEMT indicate the models. Subscripts h and p identify host matrix and particulate and $P_p$ is particulate volume fraction. MGT and BEMT equations are the same to first order in volume fraction, $P_p$. MGT and ScEMT differ in first and second order of volume fraction.

$$\frac{\mu_{MGT}}{\mu_h} = 1. + 3\frac{\mu_p - \mu_h}{\mu_p + 2\mu_h}P_p + 3\frac{(\mu_p - \mu_h)^2}{(\mu_p + 2\mu_h)^2}P_p^2 \ldots\ldots \quad (7)$$

$$\frac{\mu_{BEMT}}{\mu_h} = 1. + 3\frac{\mu_p - \mu_h}{\mu_p + 2\mu_h}P_p + 9\mu_p\frac{(\mu_p - \mu_h)^2}{(\mu_p + 2\mu_h)^3}P_p^2 \ldots\ldots \quad (8) \text{ and}$$

$$\frac{\mu_{ScEMT}}{\mu_h} = 1. + 3\frac{\mu_p - \mu_h}{\mu_p + A_p\mu_h}P_p + 9\mu_p\frac{(\mu_p - \mu_h)^2}{(\mu_p + A_p\mu_h)^3}P_p^2 \ldots\ldots \quad (9)$$

MGT can be derived from a single particle scattering theory, [22] or [23]. In MGT, neither clustering nor any percolation model physics are included. That suggests it should be accurate at very low volume fractions where particulates are isolated. Therefore, it is assumed that the MGT is the most accurate at low volume fractions and thus MGT is considered a baseline to which BEMT and ScEMT are compared. Measured data at fractions below about 10% verify that expectation.

A low fraction correction term for BEMT is derived by subtracting Equation 7 from 8. The correction term $B_c$ is;

$$B_c = 6\mu_p \frac{(1-\frac{\mu_h}{\mu_p})^2(1-2\frac{\mu_h}{\mu_p})}{(1+2\frac{\mu_h}{\mu_p})^3} P_p^2 + \cdots \qquad (10)$$

Most of composites that are modeled in this paper contain large susceptibility particulates. The large permeability solution for BEMT reduces to approximately $B_c = 6P_p^2$ for nonmagnetic hosts.

A similar subtraction is formed for ScEMT and MGT. The first and second order differences between Equations 9 and 7, are:

$$Order P_p: \frac{\mu_{ScEMT}}{\mu_h} = 3P_p \frac{\mu_p - \mu_h}{\mu_p + A_p \mu_h} \left\{ \frac{1}{\mu_p + A_p \mu_h} - \frac{1}{\mu_p + 2\mu_h} \right\} \text{ and} \qquad (11)$$

$$Order P_p^2: \frac{\mu_{ScEMT}}{\mu_h} = 3P_p^2 (\mu_p - \mu_h)^2 \left\{ \frac{3\mu_p}{(\mu_p + A_p \mu_h)^3} - \frac{1}{(\mu_p + 2\mu_h)^2} \right\} \qquad (12)$$

Figure 2 shows comparison of CMA, MGT, ScEMT and measured data for modest susceptibility iron oxide and larger susceptibility NiFe composites. As previously noted, MGT agrees well with measured data for fractions below about 10% and for lower susceptibilities. The corrected BEMT prediction approaches measurement in the same fraction range; however, at higher fractions it incorrectly predicts a sudden increase for large susceptibility particulate as the 33% percolation threshold is approached. ScEMT predictions decrease slightly and approach measurement below 10% but in general the ScEMT correction is not sufficient at low volume fractions. In addition, ScEMT predicts an artifact vs. fraction. Since $A_p$ is a function of volume fraction, the ScEMT correction term can show a sign change near 10% and thus increase rather than decrease the prediction. Such a susceptibility increase is not observed in any measured data that was available to the author. Therefore, in these discussions, the fraction value where the ScEMT correction changes sign is considered a limit for accuracy of the model. ScEMT measurement disagreement suggests that it should not be applied at low volume fractions; but electronic applications require higher volume fractions and thus its usefulness is not diminished.

When comparing model accuracy, different measurements sets of the same composite formulations are valuable. However, limited data was found related to formulation-measurement repetition. The authors of reference [24] formulated and tested composites of two NiFe-polymer composites with "identical" volume fractions. Data are shown in Figure 2b . Measurements had different particle size distributions 2.5 and < 45 $\mu m$ but both are well below the size of skin depth effects in the MHz range. Note that those measurements differ by 10 % at the duplicated fractions near 60 and 65 %. EMT should not be expected to be more accurate. Duplicate data is also seen in the upcoming Figure 4 that shows Fe particulate data. Again, measurements differ by 10-20% for the same fractional content. Similar variance is expected for composite measurements for all magnetic particulates.

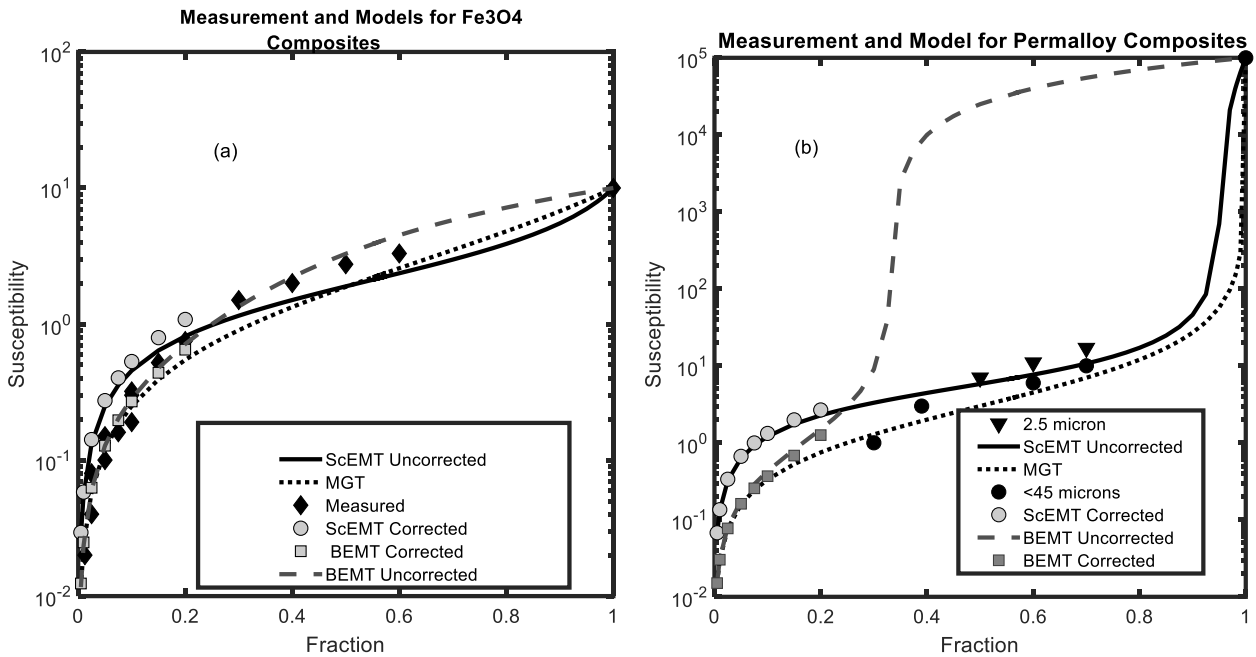

Figure 2 EMT model predictions with ScEMT corrected by Equations 11-12 and BEMT by Equation 9. (a) Left is a moderate susceptibility particulate ($Fe_3O_4$ with susceptibly 14), (b) right is for a high permeability Permalloy of susceptibility near $10^5$.

Table 1 identifies the ferrites and ferromagnetic particulates for which composite data were found in publications. Some data sets (e.g. [28]) were incomplete but valuable for their references. The majority of publications included information on bulk susceptibility, resonant, relaxation frequencies and composite microstructure. Figure 2, Figure 3 and Figure 4 show examples of ScEMT predicted and measured susceptibility from recent sources and are additions to those shown in [4]. These more recent ScEMT susceptibility predictions continue to show better agreement with measurement than BEMT, MGT and CMA models at mid volume fractions (10 – 100%) and large susceptibility particulates.

In the following discussions the relaxation frequency is taken as the frequency difference between upper and lower frequencies at the ½ magnitude of the imaginary part of the susceptibility. Resonant frequency is taken as that frequency where the imaginary permeability peaks. In most cases, particulate resonant and relaxation frequency were derived from graphs found in the references. In some cases, the graphical data could be compared and verified by tabulations, e.g. Lax and Button [10]. However, the reader should assume that there is some level of author reading error that is embedded in the data of Table 1. Therefore, the derived data in Table 1 should not be used as "absolutes" but are representative values that can be used for model comparisons. A full validation of models must await a series of carefully controlled experiments similar to those summarized in the conclusion. The following sections utilizes composite susceptibility and the tabulated bulk properties in prediction of resonant frequency and relaxation frequency.

Table 1 Table cross referencing reference source and magnetic particulate parameters

| **Particle Composition** | **DC susceptibility** | **Resonant Frequency (GHz)** | **Relaxation Frequency (GHz)** | **Reference(s)** |
|---|---|---|---|---|
| NiZn, MnZn Ferrite industrial powders | ~ 3000 | unknown | unknown | [28] |
| $Mn_{.53} Zn_{.41} Fe_{2.06} O_4$ | ~ 4000 | .0008 - .001 | .003 - .004 | [9], [29] |
| $Ni_{.24} Zn_{.65} Fe_{2.04} O_4$ | 1470 | .0015-.0025 | .017 | [9], [29] |
| $Fe_3O_4$ | ~ 14 | 1.3 | 1.6 | [30], [31] |
| $Ni Mn_{.02} Fe_{1.9} O_4$ | 27.3 | .1 | .8 | [32] |
| $(MnO)_{.25} (ZnO)_{.2115} Fe_2 O_3$ | 1750 | .001-.002 | .01-.02 | [33] |
| $BaCoZn Fe_{16} O_{37}$ | 11 | 1.4 | 1.7 | [34] |
| NiFe, e.g. $Fe_{53}Ni_{47}$ | 200-800 | .008 | ~ .016 | [35], [40], [24] |
| Fe | ~ 40 | 1.3 | 1.5 | [36]-[39] |
| $Ni_{.36} Zn_{.64} Fe_2 O_4$ | 1200 | .007 - .009 | .02-.03 | [39] |
| $Ni_{.7} Zn_{.3} Fe_2 O_4$ | 20.4 | .62 - .70 | 1.0 | [10], [13] |
| $Ni_{.3} Zn_{.58} Cu_{.08} Fe_{2.03} O_4$ and $Ni_{.24} Zn_{.65} Cu_{.07} Fe_{2.04} O_4$ | 839 and 863 | .009 and .005 | .01 and .02 | [20], [41] |

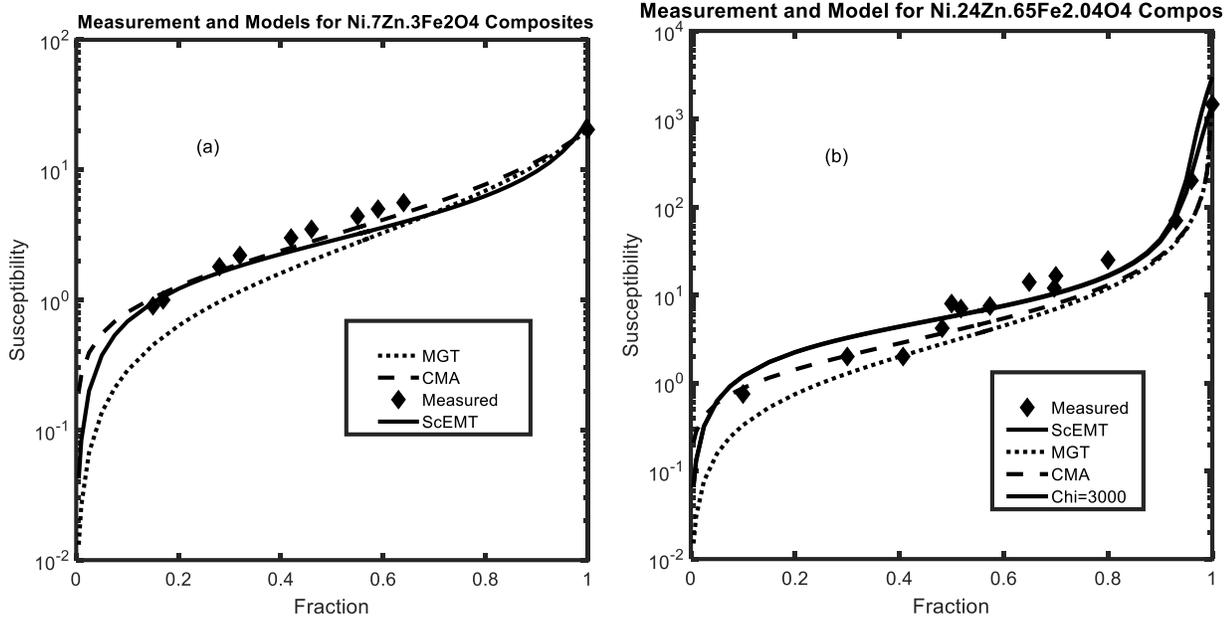

Figure 3  EMT model comparisons to measured datra for a particulate with susceptibility (a) left near 20 and (b) right 3000

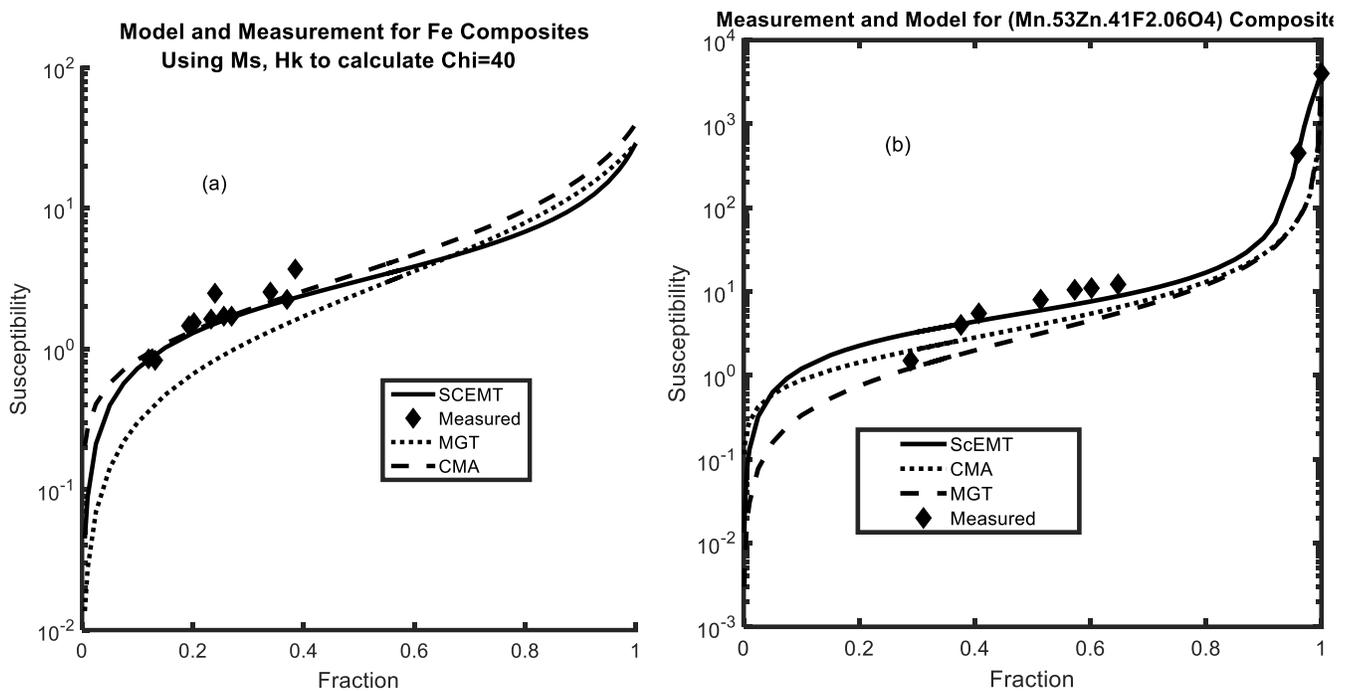

Figure 4 EMT model predictions for (a) left Fe particulate and (b) right $Mn_{.53} Zn_{.41} Fe_{2.06} O_4$ composites. Fe data were taken from a range of commercial EMC products. The $Mn_{.53} Zn_{.41} Fe_{2.06} O_4$ has a permeability near 4000.

## Characteristic Composite Frequencies

Prediction of composite susceptibility is the first step toward calculating frequency dispersive spectra of mixtures. It is assumed that bulk susceptibility, resonant and relaxation frequency will be available as input for any design studies.

The addition of a magnetic particulate to a nonmagnetic matrix produces a material whose resonance is shifted to a higher frequency than that of the particulate. Reference [9] approached the prediction of composite resonance, and relaxation frequency, by inserting an analytical equation for composite susceptibility (i.e. MGT or CMA) in the Landau-Lifshitz-Gilbert equation. The resulting dispersive models take forms that easily identify equations for composite resonant, $f_{rc}$ and relaxation frequencies, $f_{dc}$. Both $f_{rc}$ and $f_{dc}$ are proportional to particulate bulk resonance $f_{r0}$ and relaxation $f_{d0}$. MGT and CMA derived relations for the resonance and relaxation parameters are shown below. When expressed as a ratio, the relative composite resonant frequency ratio is the square root of the relaxation ratio. The equations are attractive for their simplicity and intuition. However, they are found to under predict the volume fraction induced shift in resonant frequency for many magnetic particulates.

$$CMA\ [9]: \frac{f_{rc}}{f_{r0}} = \{1. + \chi_p \left(1 - P_p^{\frac{1}{3}}\right)\}^{1/2} \quad (13)$$

$$CMA\ [9]: \frac{f_{dc}}{f_{d0}} = \{1. + \chi_p \left(1 - P_p^{\frac{1}{3}}\right)\} \quad (14)$$

And

$$MGT\ [9]: \frac{f_{rc}}{f_{r0}} = \{1. + \chi_p(1 - P_p)\}^{1/2} \quad (15)$$

$$MGT\ [9]: \frac{f_{dc}}{f_{d0}} = \{1. + \chi_p(1 - P_p)/3\} \quad (16)$$

References [25] and [26] take a different approach to determine resonance. They apply integral relationships that leverage Snoek's law [27]. A similar approach, which leverages the ScEMT predictive model for susceptibility, is investigated below and compared to Equations 13, 15 and measurement. Equation 14 and 16 will be applied in the next section on relaxation frequency.

The magnetic composites considered in this paper contain single or multidomain particulates and their mixture forms a separate material that is effectively multidomain. Therefore, a form of Snoek's law should apply for the isotropic composites that are topics of this of this paper. In Snoek's law the product of particulate bulk susceptibility, $\chi_p$ and resonant frequency $f_{rp}$ are related to particulate bulk magnetization, $M_S$, and $\gamma = 2.8\ MHz/Oe$ i.e.

$$\chi_p f_{rp} = \frac{2}{3}\gamma 4\pi M_s \qquad (17)$$

ScEMT is combined with Snoek's law by assuming that the product of composite susceptibility and resonant frequency are equal to a constant multiplied by the bulk material magnetization. We choose a magnetization that is linearly scaled for volume fraction of the particulate.

$$\chi_c f_{rc} = \gamma P_p \frac{2}{3} 4\pi M_s \qquad (18)$$

Simple algebra shows composite susceptibility, $\chi_c$, will be related to composite relaxation frequency, $f_{rc}$, bulk susceptibility, $\chi_p$, volume fraction, $P_p$, and bulk resonance frequency, $f_{rp}$, by ,

$$f_{rc} = \frac{P_p f_{rp_r}}{\chi_c}\chi_p \qquad (19)$$

Note that this approach requires a model that predicts accurate values for composite susceptibility. Since ScEMT has been found to predict susceptibility better than other EMT for large susceptibility particulates in the in the mid volume fraction range, it is not surprising that improvements are evident for other magnetic parameters like resonance frequency. Examples for EMT model resonant frequency predictions are shown in Figure 5, Figure 6, Figure 7 and Figure 8. Predictions and measurement are plotted relative to the particulate bulk value.

The trends illustrated in this figure collection are representative of the twelve composites that have been evaluated to date. In general, ScEMT is most accurate in the prediction of resonant frequency for composites using a large susceptibility pigment. As volume fraction decreases CMA, MGT and ScEMT all deviate from measurement. However, this may partially be an artifact due to the broadening and decreased magnitude of the resonance and difficulty in reading precise values. The NiZnCuFerrite plot (Figure 5b) is the only one in which digital data were available and as was previously noted, most data were acquired by reading values from plots. Thus, data should not be taken as absolutes. Even though susceptibility predictions were close, Figure 4, the resonant frequency data for Fe composites were the poorest match of the 12 materials evaluated to date, Figure 7. This may reflect the use of commercial products which use a wide range of Fe particulate compositions. For example, some products use coated particulates to change composite electrical conductivity. Overall, ScEMT is found to improve prediction of composite resonant frequency as compared to MGT and CMA.

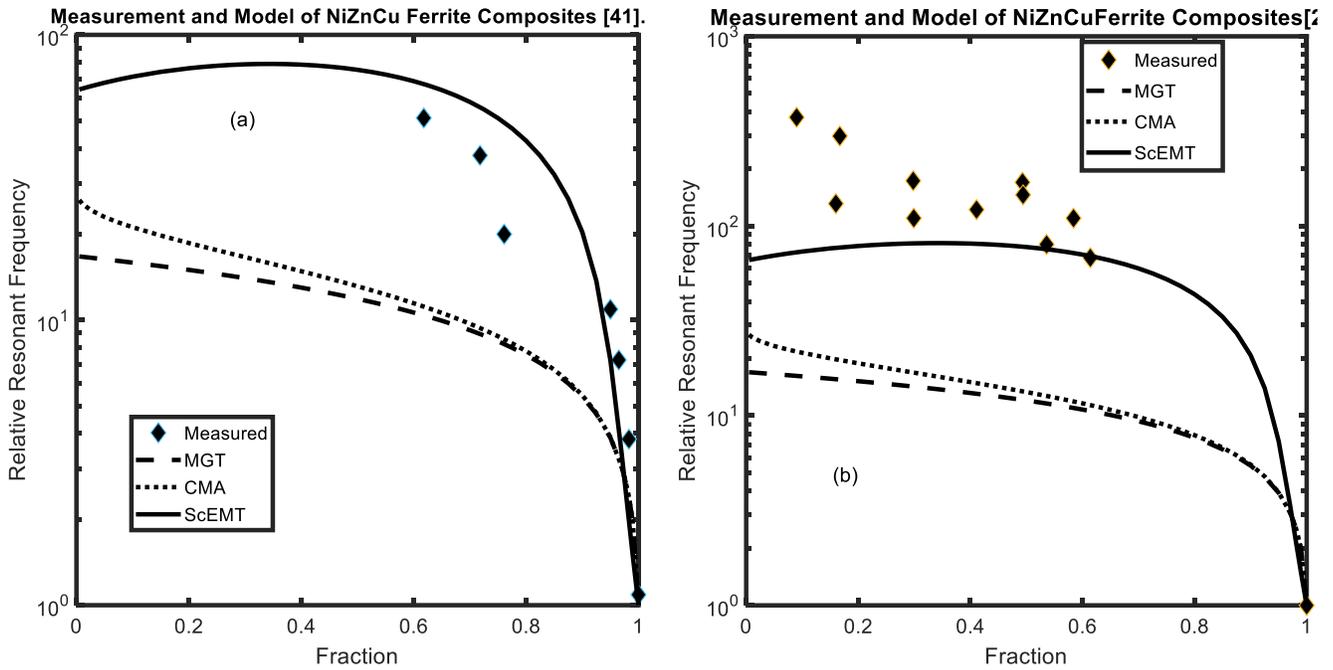

Figure 5 ScEMT, CMA and MGT model predictions for measured resonant frequency of composites. (a) left are composites of $Ni_{.3} Zn_{.58} Cu_{.08} Fe_{2.03} O_4$ with susceptibility of 839 [41] and (b) right is $Ni_{.24} Zn_{.65} Cu_{.07} Fe_{2.04} O_4$ with susceptibility 863 [20]

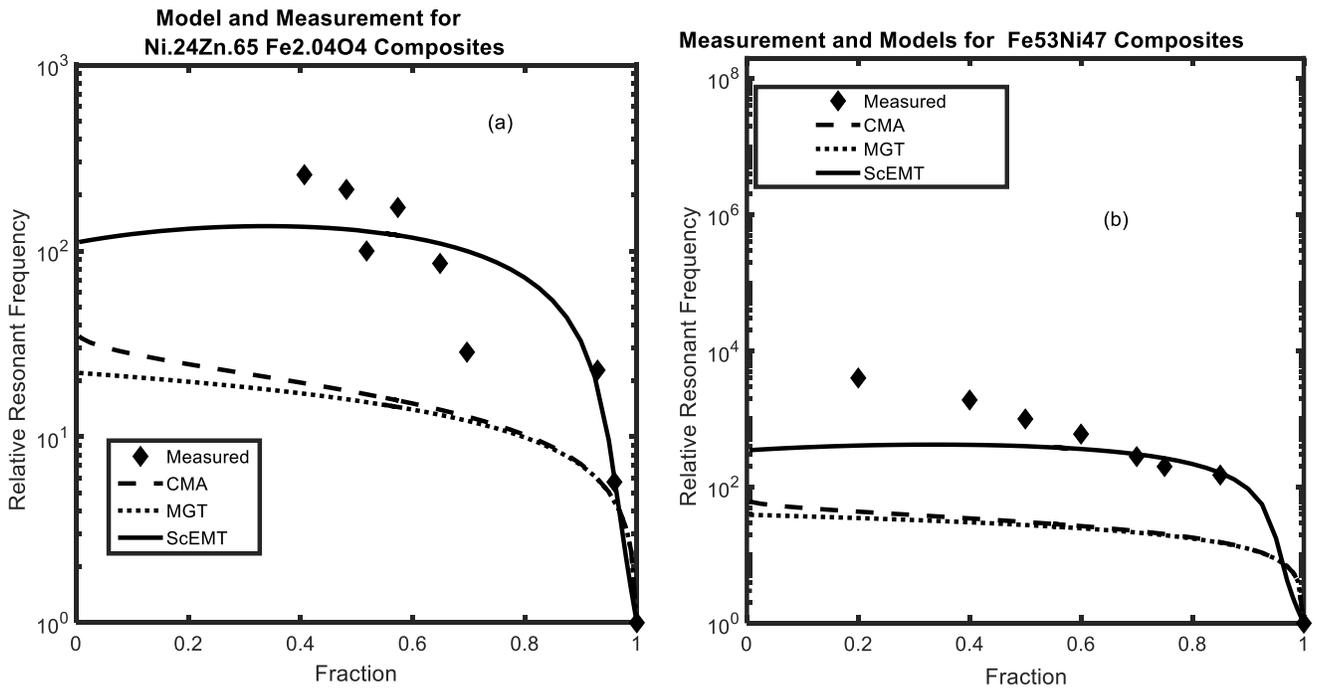

Figure 6 ScEMT, CMA and MGT model predictions compared to measured resonant frequency of composites for high susceptibility particulates: (a) left susceptibility 1467 and (b) right 800 - 4000.

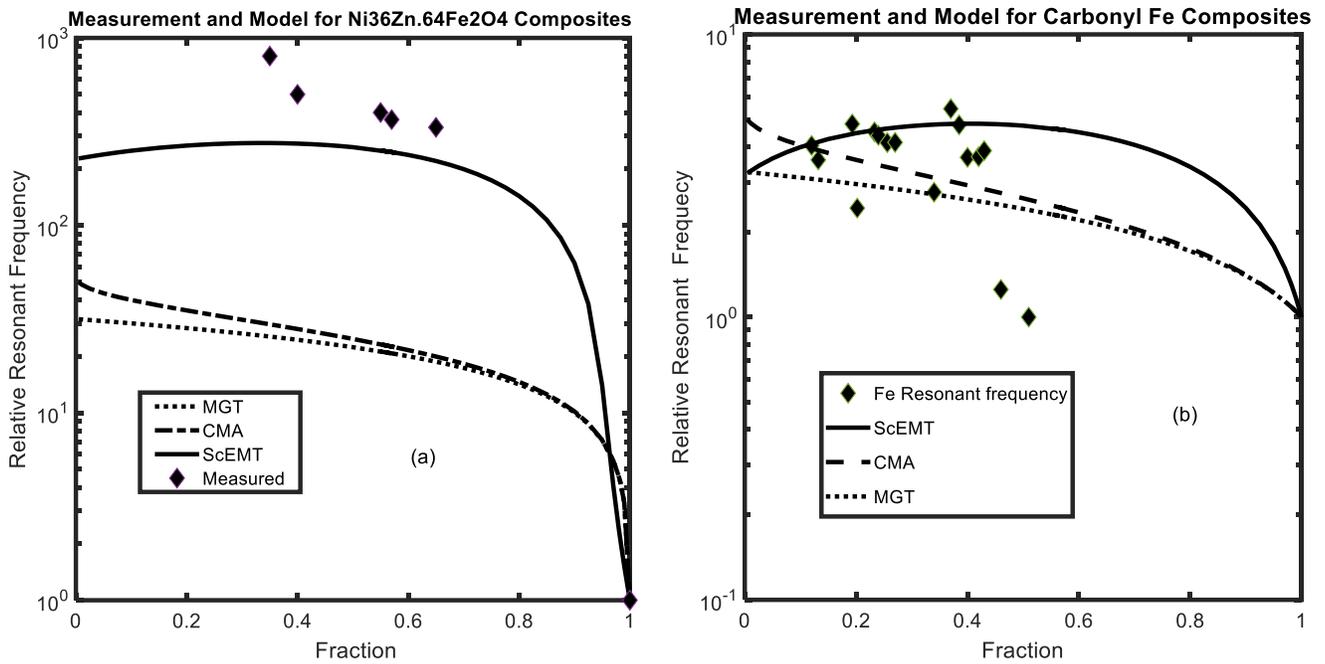

Figure 7 ScEMT, CMA and MGT model's prediction of resonant frequency vs. fraction for high and modest susceptibility particulates: (a) left 1200 and (b) right 40. Figure 7(b) shows the least favorable comparison of model and measurement for composites.

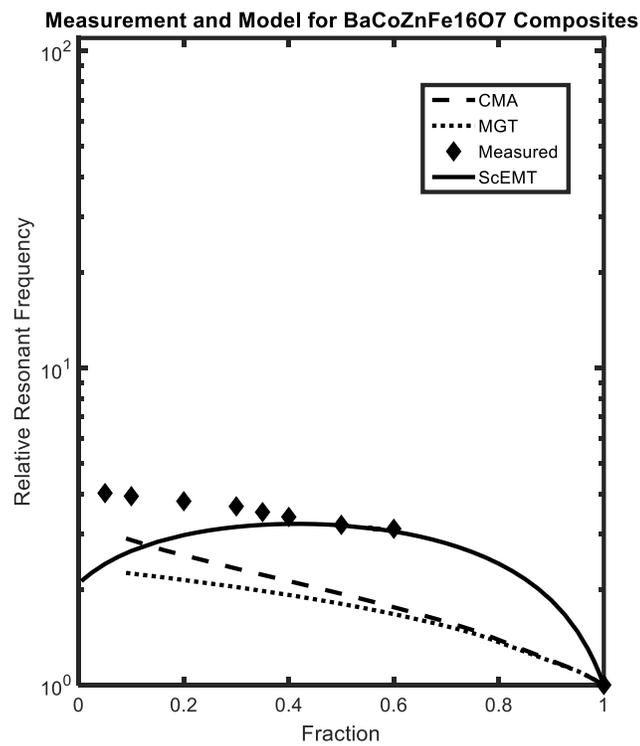

Figure 8 ScEMT, CMA and MGT model's prediction of resonant frequency vs. fraction for the BaCoZn ferrite.

**Relaxation Frequency Model and Measurement**

In 1956 E. Schlomann [7] published a composite model which approximately predicted the change in resonance bandwidth (i.e. relaxation frequency) for ferrites which contained nonmagnetic voids with fraction $P_v$ and fractions less than about 30%. Schlomann's equation from [7], is reproduced in Equations 20. Equation 20 is written in terms of void volume fraction ($P_v$) and 20b as a function of magnetic material volume fraction, ($P_p$)

$$\Delta f_{dc} = \gamma \left[\frac{K_1}{M}\right] + 1.5\gamma(4\pi M_s)(\frac{P_v}{1+P_v}) \qquad (20a)$$

$$\Delta f_{dc} = \gamma \left[\frac{K_1}{M}\right] + 1.5\gamma(4\pi M_s)(\frac{1-P_p}{2-P_p}) \qquad (20b)$$

$K_1$ is the first order anisotropy constant; M is magnetization, $\gamma$ is 2.8 MHz/Oe. The term $\frac{K_1}{M}$ represents the contribution from crystalline anisotropy and is equal to $0.5 H_k$ where $H_k$ is the anisotropy field. The second term represents the internal induced fields due to magnetic poles that are produced at the interfaces of magnetic and nonmagnetic material and is a function of magnetization and fraction. The volume fraction dependence shown in 20a and 20b were derived for a simple spherical magnetic particle with a center located void and the equations assumes a small volume fraction of void. Note that Equations 20 a, b represent the *additional* resonance broadening due to the nature of the mixture. In order to obtain the overall relaxation frequency, the original bulk value for relaxation frequency would be added to Equation 20a or 20b.

In 1969 Schlomann published a more extensive analysis [8] that changed the lead term of Equations 20 for magnetic materials with anisotropy fields smaller than $4\pi M_s$ (i.e. large susceptibility), Large susceptibility ferrites are the emphasis of this paper and thus Equation 21 (from [8] and [41]) will be the choice for modification in this article. Equation 21 was compared to measurement and performed well for ferrites containing void concentrations less than 30%, [8] Obviously the model must be modified for application to the entire volume fraction range from 0 to 100 %. This article seeks to expand application and thus the volumetric scale of Equation 20b is replaced by a "volume scale function to be determined" $D(P_p)$.

$$\Delta f_{dc} = \gamma 2.07 \left[\frac{H_k^2}{4\pi M_s}\right] + 1.5\gamma(4\pi M_s)D(P_p) \qquad (21)$$

Not only are Equations 20 and 21 based on the simple model of spherical void inside a spherical volume. But some numerical values may be approximate. For example, the multiplier 1.5 in the second term derives from an approximate contour integral [7]. Subsequent experimental research by Pointon and Roberson [42], Gurevich, et. al. [43] and R. Krishnan [44] suggested modifications to this constant. The authors measured both spinel and garnet ferrites which contained pores and they found that the constant 1.5 should be replaced by a parameter (here called $C$) and that parameter should be close to unity.

With that background, the development of a modified model is begun by inserting Equation 17 into the lead term of Equation 21. This produces an anisotropy contribution of $\Delta f_{dk}$ which is a function of ferrite susceptibility and ferrite resonant frequency,

$$\Delta f_{dk} = 1.38 \frac{f_r}{\chi_P} . \quad (22)$$

The second term of Equations 20 or 21 is a function of magnetization and volume fraction. It is first multiplied by unity in form of (3/2)(2/3) and Equation 17 is again inserted to rewrite the magnetization contribution in terms of bulk resonant frequency, $f_r$, and particle susceptibility $\chi_p$,

$$\Delta f_{dms} = 1.5 \gamma 4\pi M_s D(P_p) = 1.5 \frac{3}{2} D(P_p) \left( \frac{2}{3} \gamma 4\pi M_s \right) = C D(P_p) \chi_p f_r \quad (23)$$

Numerical factors have been collected into the parameter $C$.

Equations 22 and 23 are summed to give the composite relaxation frequency, $f_{dcs}$ as

$$f_{dcs} = \Delta f_0 + \Delta f_{dk} + \Delta f_{dms} = \Delta f_0 + \chi_p f_r \left\{ \frac{1.38}{\chi_P^2} + C D(P_p) \right\}. \quad (24)$$

The second term of Equation 24 must be modified to include a volume scale which is appropriate to a full range of volume fractions. Choices for this function were developed in [3].

As stated in [7], the second magnetization term contribution originates from inclusions which give rise to free magnetic poles inside the ferrite composite and thus to an additional magnetic field which contributes to demagnetization. Insertion of magnetic particles in a nonmagnetic matrix produces the same effect. The derivation of the scaled demagnetization coefficient of the effective medium theories (ScEMT) should be proportional to the density of poles and it is tested as one volumetric scaling function $D(P_p)$ of Equation 24.

That demagnetization coefficient, $A_p$, was developed in [3] and repeated in Equation 25.

$$A_c = A_0 \{ 1. - A_1 (1 - P_p)^\gamma \} \text{ and } A_p = \left( \frac{1 - A_c}{A_c} \right) \quad (25)$$

As noted in the previous ScEMT Review section, the values for the ScEMT scaling function parameters were determined from a fit to measured values of one ferrite-composite susceptibility. The same coefficients have been applied to all other composites. The derived parameter values are: $A_0 = 0.975$, $A_1 = 0.923$ and $\gamma = 1.210$, when rounded to the nearest thousandth. Coefficients are nearly equal but do not quite meet boundary conditions for ScEMT and BEMT equations. If the coefficients were exactly equal; then $\left(\frac{1-A_C}{A_C}\right)$ is zero at unity volume fraction and ScEMT would predict particulate-composite equality at unity volume fraction.

The fitted parameters in the ScEMT model have this ratio approaching 0.025 as concentration goes to unity. At zero concentration the Equation 25 ratio approaches 12.32 whereas it should go to infinity. These small numeric values indicate that small errors should be expected at high and low volume fractions. In the following calculations the parameter values are equalized ($A_0 = A_1 = 1.0$) thus satisfying boundary conditions. In addition, the numeric constant $C$ is set equal to unity. Figures will compare Schlomann, CMA, MGT and Equation 26 with measurement and the scaled from of Schlomann is referred to as "Scale Eq. 26, unity".

$$f_{dcs} = \Delta f_0 + \Delta f_{dk} + \Delta f_{dms} = \Delta f_0 + \chi_p f_r \left\{\frac{1.38}{\chi_P^2} + CA_p\right\}. \qquad 26$$

Reference [3] also described a second scale for demagnetization that was an Anzatz designed from micromagnetic simulations of mixtures of particulates with high susceptibility. The volume scale was of the form $\left\{\frac{1-P_p}{P_p}\right\}^x$. Physically this represents the ratio of volume fractions for nonmagnetic to magnetic material. Simulations indicated that the power, $x$, was near 3/2. In the following we apply the scaling form but set x=1. Predictions using this scaling are referred to in the figures as "Volume Ratio".

Overall, model and measurement agreement separate into two sets. Relaxation frequencies for composites containing large susceptibility particulate were best predicted using Equation 26 with either of the two volumetric scales. Predictions are consistent with Schlomann's specification that magnetic materials should have anisotropy fields smaller than $4\pi M_s$ ( large susceptibility[8]). Graphs of model and measurement comparisons for large susceptibility particulates are shown in Figures 9-11. Particulate susceptibility assumes values that are taken from the referenced publication. Notable is data for Fe.$_{53}$ Ni.$_{47}$ composites. The susceptibility of 800 is smaller than expected but is derived from a fit to measurement. Values of a few hundred to a 1000 were found in references [24][40][35].

As figure numbers increase the reader will note a shift in predictive accuracy from Equation 26 to the CMA prediction and Equation 14. For example, Figures 12-13 show measurement-model comparison for composites with particulates with susceptibility in the 10 – 30 range. As with resonant frequency and susceptibility predictions, the CMA and MGT accuracy improves as particulate susceptibility decreases. Particulate coupling decreases with susceptibility and thus "isolated particle" models like CMA and MGT better represent the microstructure of the composite.

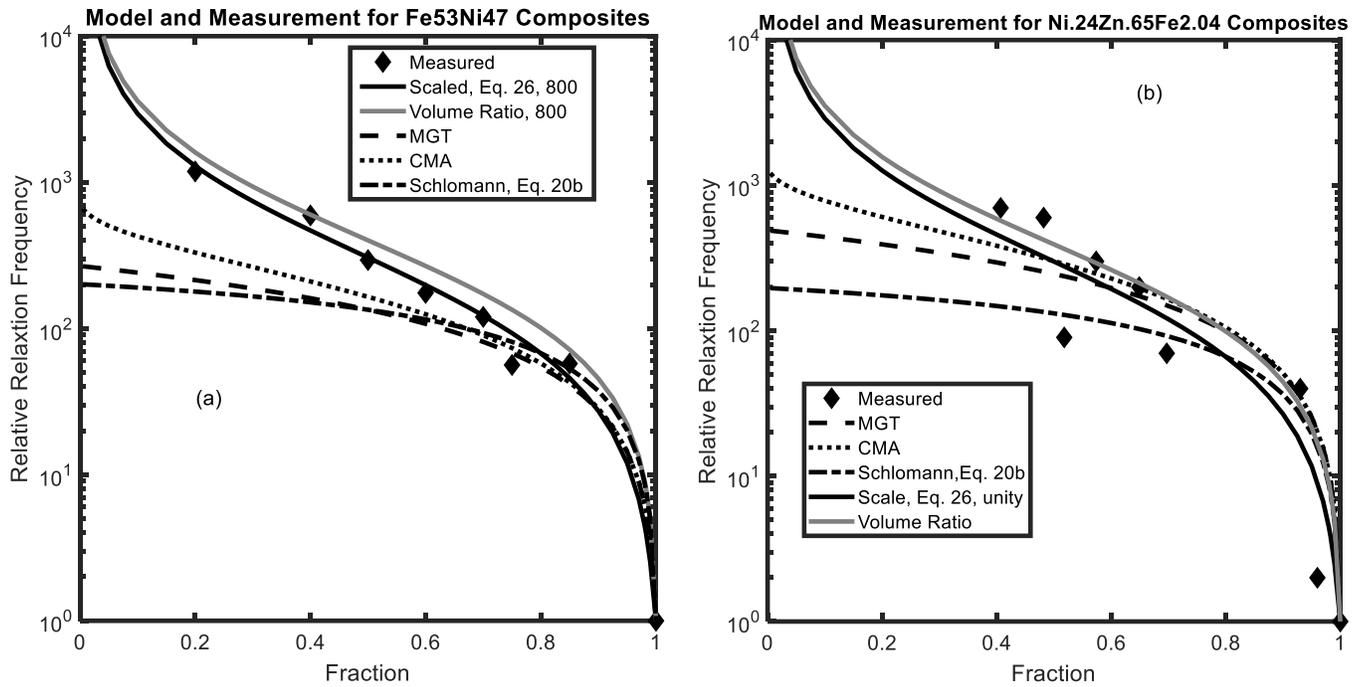

Figure 9 Measurement and EMT predicted relaxation frequency for particulate composites with large permeability. Good fits are found for Coe~1.0. Particulate susceptibilities are: (a) left 800 and (b) right 1470.

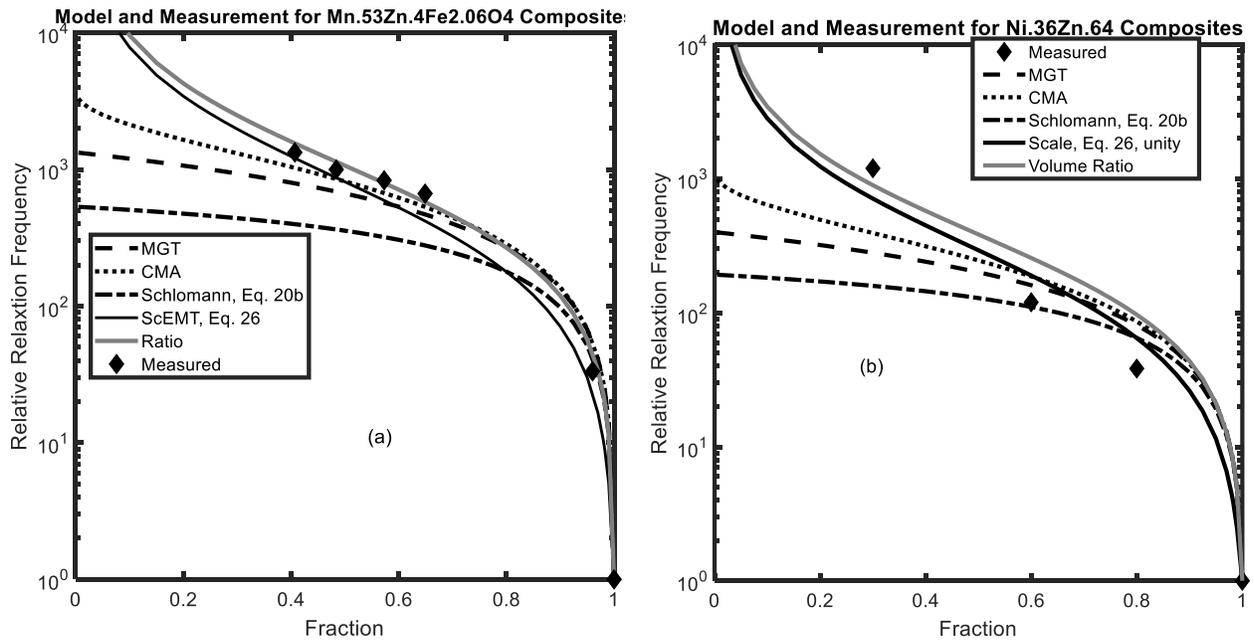

Figure 10 Measurement and EMT predicted relaxation frequencies for the indicated particulate. Particulate susceptibilities are: (a) left 4000 and (b) right 1200.

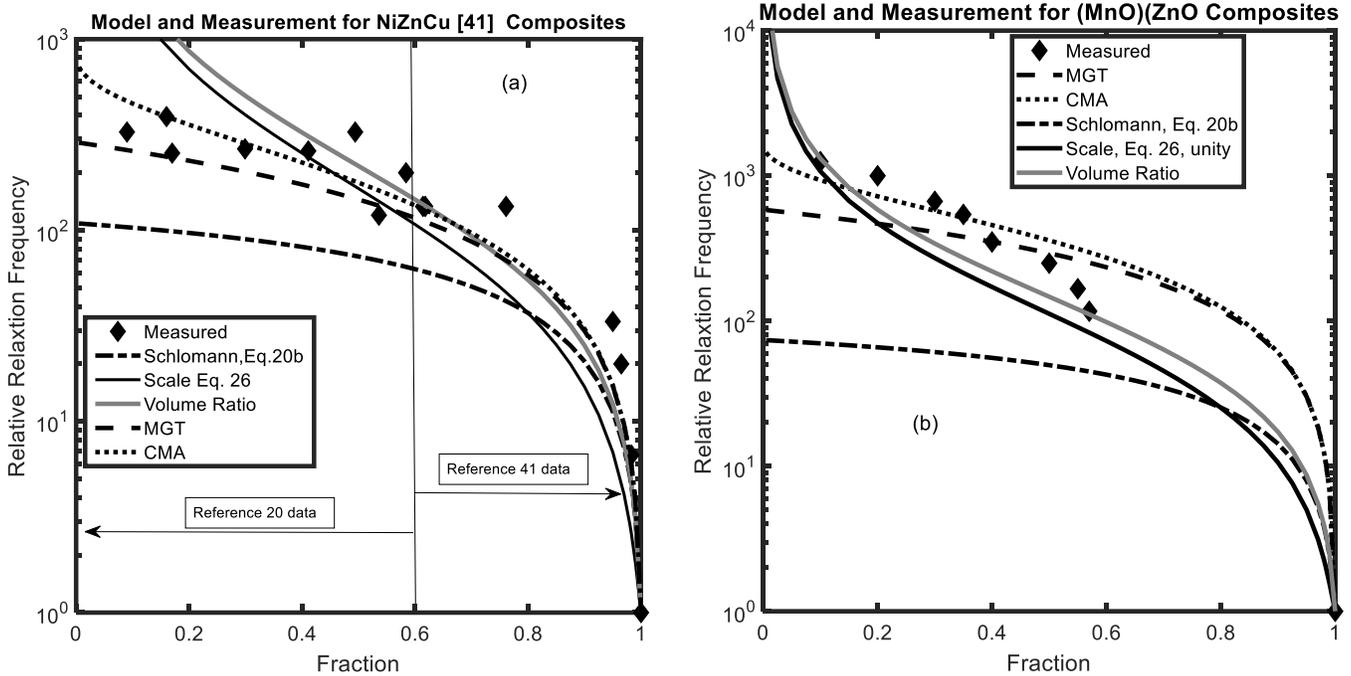

Figure 11 Measurement and EMT predicted relaxation frequencies for the indicated particulate. The NiZnCu compositions chemistries were close but measured susceptibilities differ by about 3%. Thus, two comparisons are shown. Particulate compositions and susceptibilities are: $Ni_{.3} Zn_{.58} Cu_{.08} Fe_{2.03} O_4$, 839 to the right of the vertical line [41]; $Ni_{.24} Zn_{.65} Cu_{.07} Fe_{2.04} O_4$, 863 to the left of the vertical line[20], On the right is susceptibility is 1750 or the MnO-ZnO ferrite composite.

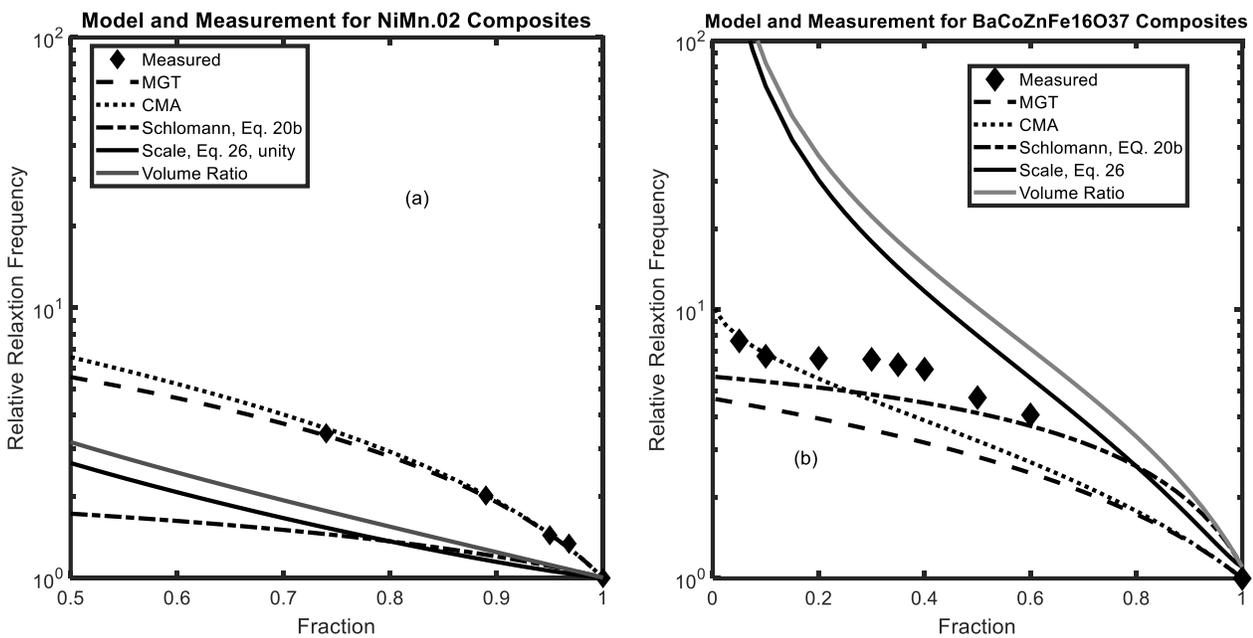

Figure 12 Measurement and EMT predicted relaxation frequencies for the indicated particulate. Particulate susceptibilities are: (a) on the left 27.3 and (b) on the right 11.0.

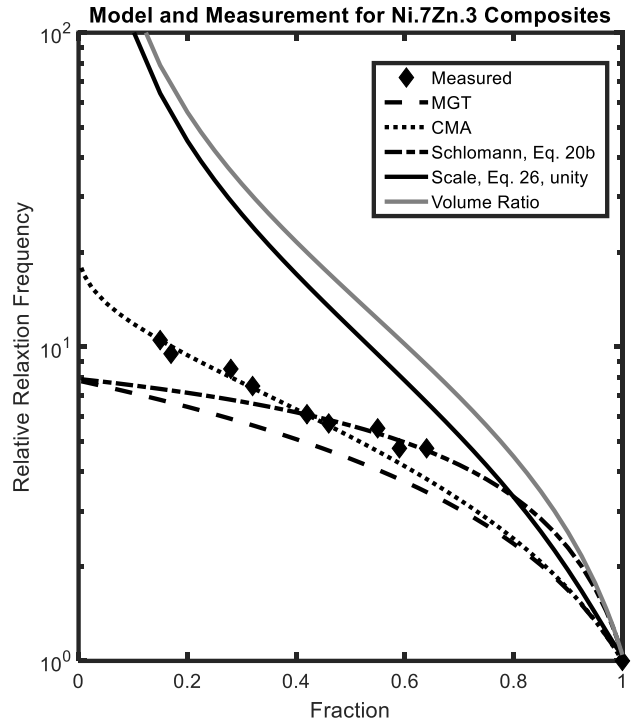

Figure 13 Measurement and EMT predicted relaxation frequencies for the indicated particulate. Particulate susceptibility is 20.4

## Conclusions and Path Forward

Maxwell-Garnett Theory (MGT), Bruggeman Effective Medium Theory (BEMT), Coherent Model Approximation (CMA), and models of Schlomann and Scaled Effective Medium Theory (ScEMT) were applied in this paper to predict magnetic susceptibility, resonant and relaxation frequency in polymer-magnetic particle composites. The paper presented a short study of modifications to BEMT and ScEMT that should correct the respective models for volume fractions below about 10%. A ScEMT based model of composite resonant frequency was presented and compared to MGT, CMA models and measurement. Model and measurement comparisons of resonant frequency were followed by model and measurement comparisons for relaxation frequency. CMA, MGT and models of Schlomann with various volumetric scales are used in the relaxation frequency study. Though ScEMT susceptibility resonant and Schlomann modified relaxation frequency are not perfect in prediction of measurement, they demonstrate significant improvement over other historical EMT in composites that use particulates with large susceptibility. Measured data for susceptibility, resonant and relaxation frequencies demonstrate a transition from isolated particulate to strongly coupled particulate composite physics. The transition is characteristic of a microstructural change from dispersive to symmetric. The better accuracy of ScEMT for large susceptibilities is supported by numerical studies [45][46]. The publications suggest that spherical particulates with large magnetization and large susceptibility, are more likely to attract one another and their attraction will exceed frictional drag of media surrounding the particles. Particles assemble into chains and large structures and therefore volume fraction scaled models are necessary to reflect the fundamental physics.

Over all, models should be supported by additional measurement. They would be applied to predict frequency dispersive susceptibility for a select set of composites and utilize bulk data for a moderate, large and very large susceptibility (e.g. 20, 800 and 4000) non-conducting ferrite. Predictions would be made for particulate sizes (1, 10 and 40 $\mu m$) and particulate volume fractions of 2, 5, 15, 20 25, 30, 35, 45 55 and 65%. Co-fired bulk samples of the three ferrites would be used as controls. Dispersive predictions will be compared to measurement over the frequency span of 10 MHz to 40 GHz.

**Acknowledgements:**

# References

[1]  Bashar Issa, et.al., Int. J. Mol. Sci. 14, 21266, (2013)

[2]  Fadzidah Mohd. Idris, et.al., J. Mag. Mag. Mat., 405, 197, (2016)

[3]  R. Moore, AIP Advances 9, 3, (8 March 2019)

[4]  R. Moore, J. Appl. Phys.,125, 8, 085101 (28 Feb 2019)

[5]  J.C. Maxwell Garnett, Phil. Trans. Royal Soc., 203A, 385, (1904)

[6]  E.G. Visser, M.T. Johnson, J. Mag. Mag. Mat., 101, 143 (1991)

[7]  E. Schlomann, Conf. on Mag. and Mag. Materials, AIEE Spec. Publ. T-91, 600 (1956)

[8]  E. Schlomann, Phys. Rev. B, 182, 7, 632 (10 June 1969)

[9]  T. Tsutaoka, et.al., 2013 IEEE EMC International Symposium Digest,.545, IEEE-978-1-4799-0409-9/13

[10] B. Lax, K. J. Button, Microwave Ferrites and Ferrimagnetics, 474-477, McGraw Hill, (1962)

[11] R. Krishnan, IEEE. Trans. Mag., MAG-7, 1, (March 1971)

[12] D.A.G. Bruggeman, Ann.Phys.24,636 (1935)

[13] A. Chevalier, M. Le Floc'h, J. Appl. Phys., 90, 7, 3462 (1 Oct. 2001)

[14] M. Anhalt, et.al., 18th Soft Magnetic Materials Conference, B-0011 (Sept. 2007)

[15] C. Alvarez, S. H. L. Klapp, Soft Mater, 8, 7480 (2012)

[16] J.L. Mattei, D. Bariou, A. Chevalier, M. Le Floc'h, J. Appl. Phys., 87, p4975 (2000)

[17] J.L. Mattei, M. Le Floc'h , J. Mag. Mag. Mat, v357 p335 (2003)

[18] J.L. Mattei, P. Laurent, A. Chevalier, M. Le Floc'h  J. Physique. IV France 7, C1-547 (1997);

[19] J.P. Clerc, G. Giraud, J.M. Laugier, J.M. Luck, Advances in Physics, 39, 3, 191-309, (1990)

[20] Rick. Moore, Electromagnetic Composites Handbook: Models, Measurement and Characterization, Chapter 12, McGraw Hill ISBN: 978-1-25-958504-3 2016T.

[21] V. Markel, J. Opt. Soc. America, 33, 7, 1244 (July 2016)

[22] I. Webman, J. Jortner, M. Cohen, Phys. Rev. B, 15, 12, 5712 (14 June 1977);

[23] L. Lewin, Radio Sci., 94, III,65 (1947)

[24] T. Kasagi, et.al., J. Mag. Soc. Of Japan, v 22, n S1 (1998)

[25] K.N. Rozanov, et.al., J.Appl.Phys., 97, 013905 (2005)

[26] K.N. Rozanov, M. Y. Koledintseva, J. Appl. Phys., 119, 073901 (2016);

[27] J.L. Snoek, Physica, 14, 207 (1948)

[28] J .J. Fiske, et.al. J.Mat.Sci., 32, 5557 (1997)

[29] T. Tsutaoka, J. Appl. Phys., 93, 5 2789 (March 2003)

[30] Zheng Hong, et.al., Chin. Phys. Lett., 26, 1, 017501 (2009)

[31] S. Liong, PhD Thesis, "A Multifunctional Approach to Development, Fabrication, and Characterization of Fe3O4 Composites" School of Material Science and Engineering, Georgia Institute of Technology, (2005);

[32] J.E. Pippin and C.L. Logan, "Initial Permeability Spectra of Ferrites and Garnets", Harvard University 1959

[33] V. Babayan, et.al., Applied Surface Science, 258, 7707 (2012)

[34] Z. W. Li, Y. B. Gan, Xu Xin, G. Q. Lin, J. Appl Phys. 103, 073901 (2008)

[35] H. Massango, T. Tsutaoka, T. Kasagi, Materials Research Express, 3, 095801 (2016)

[36] A. Medeiros Gama, et.al., J. Aerospace Tech. and Management, n1, v2, 5a (1 Jan.-April 2010)

[37] Y. Shimada, M. Yamaguchi, J. Appl. Phys., 101, 09M505 (2007)

[38] P. Ramprasad, P. Zarcher, M. Petras, M. Millr, P. Renaud, J. Appl. Phys., 96, 1, 529 (1 July 2004)

[39] J. Slama, A. Gruskova, V. Jancarik, P. Siroky, M. Lehotsky., "Modification of Frequency Spectra of Ferrite Composites", Sept. 10-12, 2007, Pilsen Czech Republic

[40] Chih-Wen Chen, Magnetism and Metallurgy of Soft Magnetic Materials, ISBN 13: 9780486649979

[41] T. Nakamura, T. Tsutaoka, K. Hatakeyama, J. Mag. Mag. Mat., 138, 319 (1994)

[42] A. J. Pointon, J. M. Robertson, Phil. Mag., 12, 725-733 (1965)

[43] A.G. Gurevich, I. E. Gubleer, A. P. Safant Evskii, Sov. Phys. Solid State, 1, 1706-1708 (1960)

[44] R. Krishnan, IEEE Trans. Magnetics, Mag-7, 202 (1 March 1971)

[45] Jorge L. C. Domingos, F. M. Peeters and W. P. Ferreira, Phys. Rev. E 96, 012603 (2017)

[46] Jorge L. C. Domingos, F. M. Peeters and W. P. Ferreira, PloS One 13 (4), e0195552 (2018).